\documentclass[journal]{IEEEtran}
\ifCLASSINFOpdf
  \usepackage[pdftex]{graphicx}
  \usepackage[font=normalsize]{caption}
  \usepackage{hyperref}
  \hypersetup{
        colorlinks=true,
        linkcolor=black,
        filecolor=black,      
        urlcolor=black,
        citecolor=black,
        pdfpagemode=FullScreen,
        }

  \usepackage{amsfonts}
  \usepackage{amsmath}

\else
\fi

\begin{document}
%
\title{Efficient~Non-Uniform~Structured~Mesh~Generation Algorithm~for~Computational~Electromagnetics}
%
%
%

\author{Apostolos~Spanakis-Misirlis

\thanks{Apostolos Spanakis-Misirlis (e-mail: \href{mailto:0xcoto@protonmail.com}{0xcoto@protonmail.com}) is with the Department of Informatics, University of Piraeus, Greece.}
\thanks{Manuscript received MONTH DD, YYYY; revised MONTH DD, YYYY.}}

%
%

\markboth{Journal of \LaTeX\ Class Files,~Vol.~14, No.~8, August~2015}%
{Shell \MakeLowercase{\textit{et al.}}: Bare Demo of IEEEtran.cls for IEEE Journals}
%



\maketitle

\begin{abstract}
Despite the rapidly evolving field of computational electromagnetics, few open-source tools have managed to tackle the problem of automatic mesh generation for properly discretizing the problem of interest into a finite set of elements (cells). While several mesh generation algorithms have been established in the field of computational physics, the vast majority of such tools are targeted solely towards tetrahedral mesh formation, with the intended primary application being the finite element method. In this work, a computationally efficient non-uniform structured (rectilinear) mesh generation algorithm for electromagnetic simulations is presented. We examine the speed, performance and adaptability against previous work, and we evaluate its robustness against a complex geometry case with a commercially-generated grid. The mesh and simulation results produced using the generated grids of the proposed method are found to be in solid agreement.
\end{abstract}


%
\IEEEpeerreviewmaketitle

\section{Introduction}
%
%
%
%
\IEEEPARstart{T}{he} field of computational electromagnetics is a rapidly developing area, with many applications revolving around the design of electronics, RF networks, antennas, propagation models, and more. 
The role of computational electromagnetics is to tackle electromagnetic problems that are too complex to be solved using analytical solutions. To achieve this, the geometry of a model is fed to a solver engine, which attempts to solve Maxwell's equations across the entire model, observing the electromagnetic response of the system \cite{davidson2010}. However, because geometries of electromagnetic devices can take all sorts of complex shapes and forms, and more importantly, because it is simply impossible to solve Maxwell's equations on an infinite number of geometry points, the input geometry must undergo discretization. This way, the number of operations becomes finite, and the problem becomes computationally feasible to tackle.

While there are various numerical methods to tackle such electromagnetic problems, the most popular techniques include the finite-difference time-domain (FDTD) \cite{yee1997}, the method of moments (MoM) \cite{mom}, and the finite element method (FEM) \cite{fem}. All of these techniques come with a variety of advantages and disadvantages \cite{solver_comparison}, but time-domain methods like FDTD are established to be particularly robust when dealing with problems in the field of microwave engineering. This is because unlike frequency-domain methods like MoM and FEM, FDTD's theory of operation is based on stimulating a broadband excitation signal, and observing the system's response across the entire frequency range of interest. This is partially achieved using Fourier transforms, converting time-domain signals to frequency-domain spectra.

However, despite FDTD's increasing popularity over the last few decades, few studies have attempted to tackle the problem of geometry discretization using open-source tools. The most notable of which is AEG Mesher \cite{aegmesher}, which proposes a method of generating three-dimensional rectilinear grids from unstructured tetrahedral meshes, produced using tools like \texttt{Gmsh} \cite{gmsh}. Despite the novelty of AEG Mesher's technique however, we identify a set of drawbacks that unfortunately introduce certain limitations we aim to address in this paper.

Namely, the grid generation algorithm is fairly slow. This is because the formation of an unstructured tetrahedral mesh (which is a computationally expensive task) is a prerequisite. Additionally, the employed algorithm appears to include processes involving lots of computations, and the rectilinear mesh generation code itself could potentially benefit from better CPU utilization using multiprocessing.

Furthermore, a major restriction AEG Mesher imposes on geometry discretization is the lack of non-uniform grid generation. While a proof of concept has been demonstrated on the original study, it is yet to be fully integrated into the package for standard applications. This is arguably among the most critical limitations of the aforementioned package, as restricting a model to uniform cuboid cells can yield an unnecessarily massive grid for the FDTD engine to deal with, ultimately leading to suboptimal solver performance.

In this work, we present an open-source non-uniform rectilinear grid generation tool, that is easily applicable to solvers like FDTD, and is up to hundreds of times faster compared to previous work.\footnote{A fighter jet model (\url{https://github.com/flintoftid/aegmesher/blob/master/examples/jet/Jet.stl}) was used as a reference for the benchmark, yielding a meshing time of $<$0.7 s compared to AEG Mesher's 95 s. Due to the unstructured mesh-generation prerequisite, this speedup ratio is expected to further scale when the geometry is input as a set of 3D solids instead of a plain and simple tetrahedral mesh.} In Section II, we lay out the the data structures and computational considerations of the proposed algorithm, and proceed to describe how the parameters of the code affect the produced grid (Section III). Section IV highlights the implementation of a tree-traversal algorithm to assist in the CAD preprocessing segment of the process. The results of an example case are presented in Section V, where the accuracy is evaluated against commercially-generated grids. After concluding and summarizing our work (Section VI), further enhancements involving mesh classification techniques are proposed as future work (Section VII).

\section{Computational Considerations}

In order for the proposed meshing technique to be computationally efficient, the employed algorithm shall make use of appropriate data structures and computational operations, such that the time and space complexity is minimized. This is particularly important for large and complex models consisting of numerous vertices, demanding more operations to process and refine accordingly.

Due to its simple syntax and modularity, the programming language of choice for the implementation of the meshing algorithm is Python. Not only does Python significantly accelerate the development process---enabling quick tests and performance evaluations---but it also provides a seamless integration with the \texttt{CadQuery} Python package \cite{cadquery}. This library simplifies CAD preprocessing, supporting the input requirements of the core segment of the meshing algorithm.

\subsection{Data Structures}

Despite its simple syntax, Python's simplicity comes with a major drawback. When it comes to lists, each item's type (\textit{boolean}, \textit{integer}, \textit{float}, \textit{string}, etc.) can be arbitrary, making lists a heterogeneous data structure. While this introduces great flexibility, it may also greatly degrade the computational efficiency of a function, as each list element constructs a separate Python object. This differs from low-level languages like C, where the elements are restricted to the predefined type of the array.

In order to tackle this issue, we employ the NumPy package \cite{numpy}, which is a library designed for high-performance scientific computing, and is particularly robust with the introduction of the \texttt{ndarray} object. Unlike Python lists, the size of NumPy arrays is static, consisting of items of the same data type. Furthermore, many operations (provided as a wrapper for low-level code) are carried out by pre-compiled code, imitating the speedy functionality of low-level languages.

\subsection{Further Speedup Optimization Attempts}

Further attempts were carried out to attempt to minimize the array operations, such that mesh generation of even complex geometries could be concluded in milliseconds rather than seconds. Various techniques were tested, including the efficient use of threads, \texttt{Cython} (designed to translate Python source code into optimized and compiled C/C\texttt{++} code) \cite{cython}, as well as \texttt{Numba}: a just-in-time compiler for accelerating Python and NumPy by converting the source to machine code \cite{numba}.

However, neither of the aforementioned attempts were found to be sufficiently easy to adapt the original code to, and were thus not implemented. In 2/3 cases, implementations were successfully set up to run, but no meaningful improvement was observed. Considering the vast majority of the computing time is spent by the solver, efforts to bring the meshing time further down were discontinued, as the initial algorithmic approach was already fast.

\begin{figure}
  \begin{center}
  \includegraphics[width=0.48\textwidth]{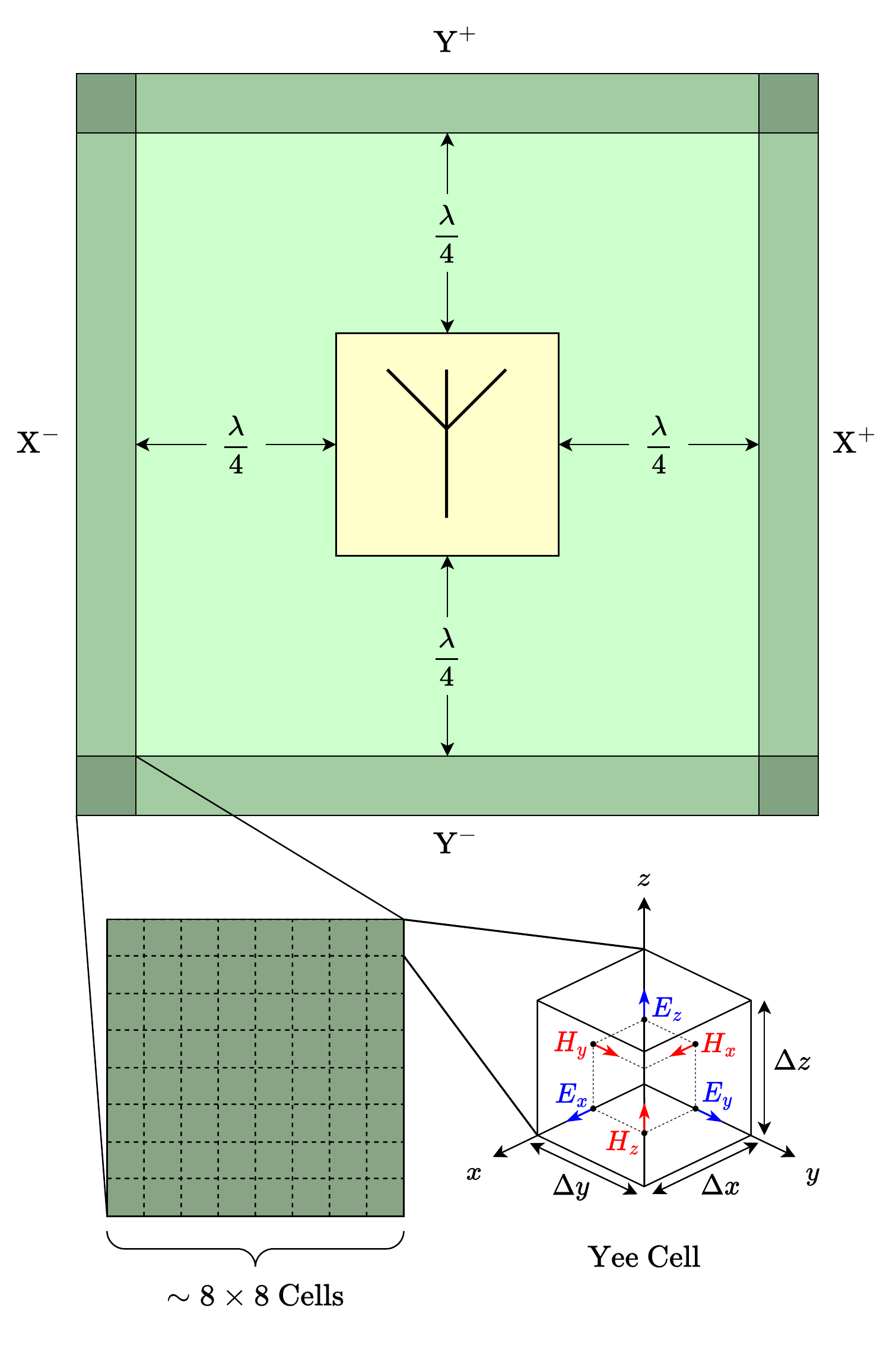}
  \caption{\textit{Top:} Two-dimensional view of the geometry placed the simulation domain. The PML zone (meant to absorb incident energy) is shaded, and is separated from the edges of the model by a quarter of the wavelength. \textit{Bottom left:} Local grid structure of the PML region, highlighting its composition of multiple cells. \textit{Bottom right:} The three-dimensional Yee cell structure of each mesh element, upon which the fundamental operation of the FDTD method is based.}\label{fig:pml}
  \end{center}
\end{figure}

\begin{figure*}
  \begin{center}
  \includegraphics[width=\textwidth]{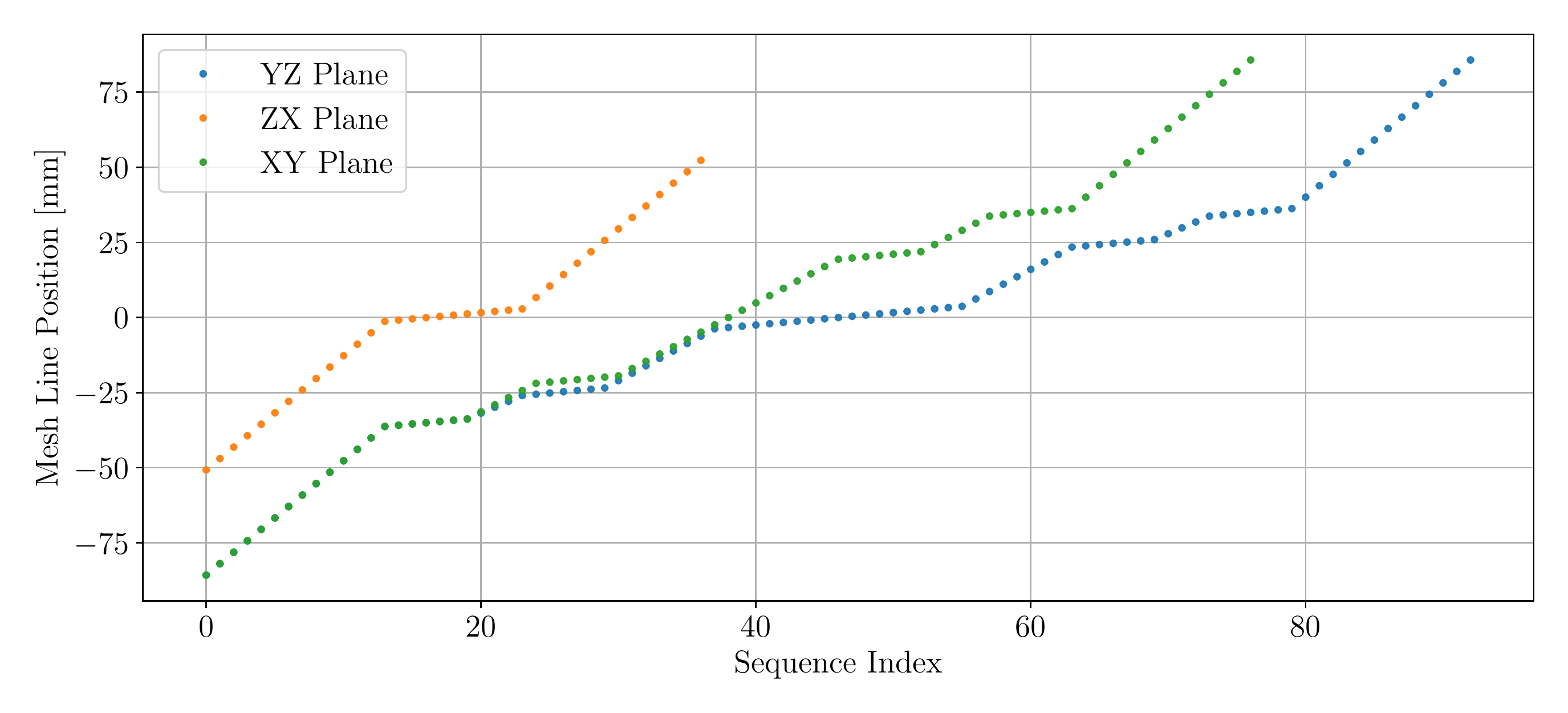}
  \caption{Mesh sequence for each plane in $\mathbb{R}^3$, consisting of $93\times37\times77=264{,}957$ mesh cells in total. Note the decreasing slope around every $m$, as the concentration of mesh lines becomes denser wherever refinement is deemed necessary by the algorithm.}\label{fig:sequence}
  \end{center}
\end{figure*}

\section{Algorithm Inputs}
In order for the algorithm to properly generate a grid suitable for the model, three inputs are required:

\begin{enumerate}
   \item[{1.}] \texttt{max\_cell\_model} \ \ \ \textit{(float)}
   \item[{2.}] \texttt{max\_cell\_space} \ \ \  \textit{(float)}
   \item[{3.}] \texttt{min\_cell\_global}  \ \textit{(float)}\\
\end{enumerate} Additionally, two optional parameters can assist with further refinements of the model for more advanced control:

\begin{enumerate}
   \item[{4.}] \texttt{n} \ \ \ \ \ \ \ \ \ \ \ \ \ \ \ \ \ \ \ \ \ \ \ \ \ \textit{(list)}
   \item[{5.}] \texttt{res\_fraction} \ \ \ \ \ \ \textit{(list)}\\
\end{enumerate} Inputs 1--3 are expressed as fractions of the shortest wavelength $\lambda_\mathrm{min}$ the excitation signal consists of. In time-domain solvers, this is determined by the maximum frequency of the Fourier transform of the excitation signal. Thus, if a generated grid is sufficiently fine for the highest frequency of the excitation, it is inherently applicable to all lower frequencies, given

\begin{equation}\label{eq:wavelength}
\frac{c}{\lambda_\mathrm{min}} > \frac{c}{\lambda} > \frac{c}{\lambda_\mathrm{max}}, \ \ \ \forall \lambda \in (\lambda_\mathrm{min}, \lambda_\mathrm{max})\\
\end{equation} where $c$ is the speed of light. However, although the level of detail may suffice, it is important to consider the space between the end of the model to the boundaries of the simulation, which are partially dependant on the longest wavelength of the excitation.

The distance between the edge of the geometry to the absorbing boundary condition (ABC) is generally given by the midpoint of the two wavelengths, $\lambda_\mathrm{mid}$, and is typically set to a quarter of the wavelength:

\begin{equation}\label{eq:wavelength_mid}
\frac{1}{4} \lambda_\mathrm{mid} = \frac{\lambda_\mathrm{min}\cdot \lambda_\mathrm{max}}{2(\lambda_\mathrm{min}+\lambda_\mathrm{max})}. \\
\end{equation} This ensures enough cells are provided for the fields to form to an acceptable degree, before reaching the ABC (and go through e.g., nearfield-to-farfield transformations).

While it is theoretically more appropriate to use a quarter of the longest wavelength $\lambda_\mathrm{max}$ instead, this becomes highly impractical when the excitation consists of very low frequencies, as the number of mesh cells can grow tremendously due to the additional space that needs to be meshed. Moreover, the simulation domain would become infinitely large if the excitation signal included a frequency $f$ equal to $0 \mathrm{\ Hz}$ (i.e., not be DC--free):

\begin{equation}\label{eq:infinite_bounds}
\lim_{f \to 0^{+}} \frac{1}{4} \lambda_\mathrm{max} = \infty.\\
\end{equation}

\newcommand{\minus}{\scalebox{0.8}{$-$}}
\newcommand{\plus}{\scalebox{0.6}{$+$}}

Additionally, depending on the type of boundaries set for the simulation, setting a quarter of the wavelength as the distance between the end of the geometry to the simulation domain may not be appropriate. For instance, because the cells of the perfectly matched layer (PML) extend inward of the boundaries, the radiated fields from the source may not have enough space to fully form, ultimately getting absorbed by PML cells prematurely. In certain applications, such setup imperfections may potentially yield inaccurate simulation results. For that reason, $\sim8$ additional cells (optionally adjustable \texttt{pml\_n} \textit{integer}, $n\in [\![4,50]\!]$) are appended to the six ends of the quarter wavelength spacings $(\mathrm{X}^{-}, \mathrm{X}^{+}, \mathrm{Y}^{-}, \mathrm{Y}^{+}, \mathrm{Z}^{-}, \mathrm{Z}^{+})$. Fig.~\ref{fig:pml} depicts this concept in two dimensions $(\mathrm{X}, \mathrm{Y})$.

\begin{figure}
  \begin{center}
  \includegraphics[width=0.42\textwidth]{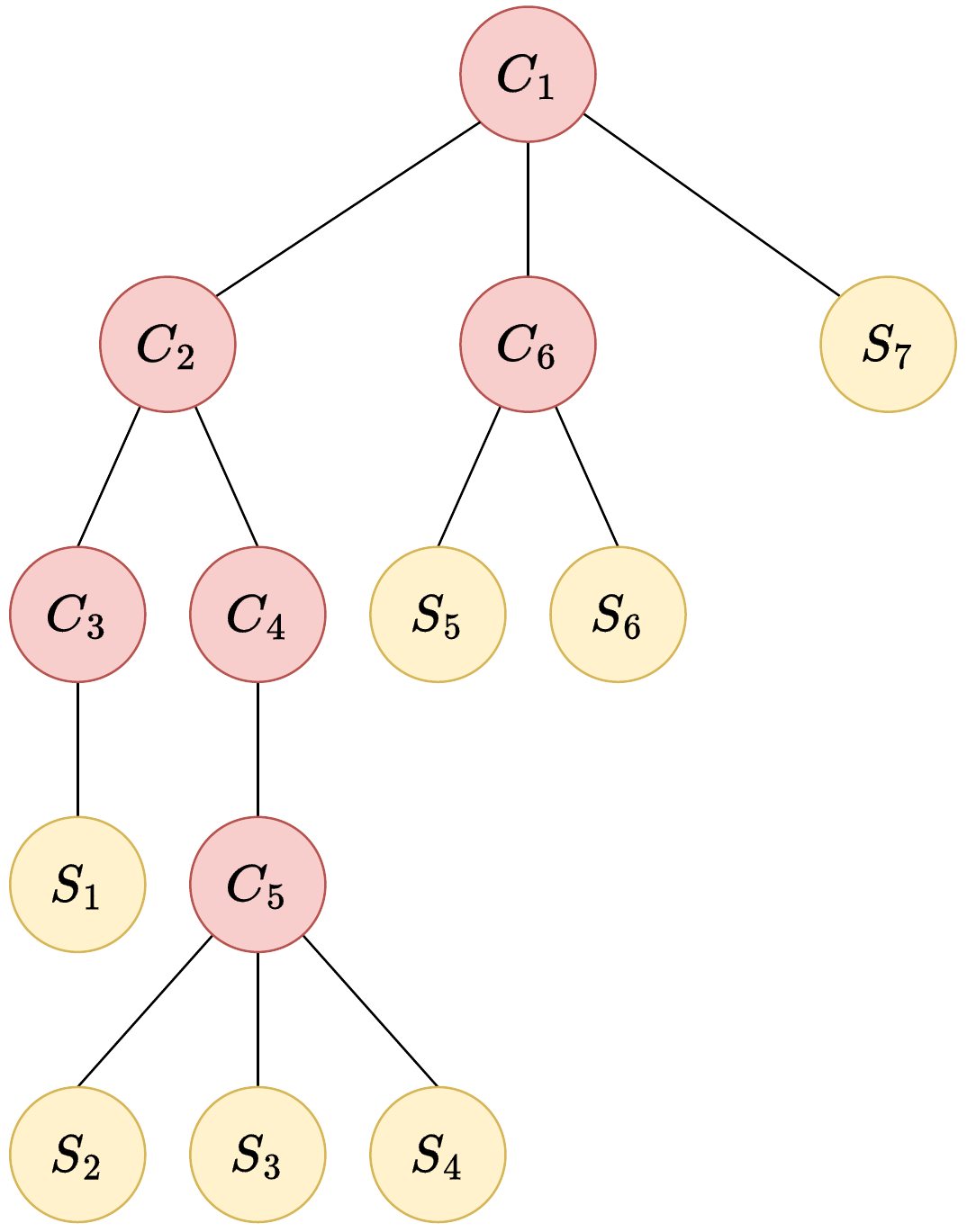}
  \caption{Tree data structure of a sample geometry model. $C_i$ and $S_j$ correspond to the $i$-th compound and $j$-th shape of the input file, respectively. Assuming the root node to be $C_1$, the output of the DFS algorithm in this case would be $C_1$,~$C_2$,~$C_3$,~$S_1$,~$C_4$,~$C_5$,~$S_2$,~$S_3$,~$S_4$,~$C_6$,~$S_5$,~$S_6$,~$S_7$. However, since we are only interested in shapes, we can eliminate all compound instances, yielding the final result of $S_1$,~$S_2$,~$\cdots$,~$S_7$.}\label{fig:dfs}
  \end{center}
\end{figure}

\subsection{Significant Parameters}

The \texttt{max\_cell\_model} parameter is the primary variable that specifies the local grid resolution of the model being simulated. The greater the value, the finer the produced mesh around the corresponding region.

Similar to this parameter, \texttt{max\_cell\_space} describes the grid resolution of the space between the ends of the geometry to the bounding box of the simulation domain. In the presence of an absorbing PML, the size of these cells are also affected accordingly. Because of the structural complexity of the model, due to abrupt differences in permittivity, $\Delta \varepsilon$, and permeability, $\Delta \mu$, the response of the electromagnetic fields tends to be highly variable and sudden. On the other hand, since the free-space region surrounding the model has a constant vacuum permittivity $\varepsilon_0$ and permeability $\mu_0$, the electromagnetic variations the waves exhibit as they propagate toward the PML zones are minimal.\\\\Thus, most cases are recommended to follow:

\begin{equation}\label{eq:model_vs_space}
\texttt{max\_cell\_model} > \texttt{max\_cell\_space}.\\
\end{equation}

Furthermore, the \texttt{min\_cell\_global} parameter determines the minimum cell dimension for all axes. Unlike the aforementioned parameters which are either only applied to the geometry, or the surrounding free-space region individually, the value of this parameter ensures no cell exceeds a specified wavelength fraction, globally across the entire simulation domain. Imposing such a constraint to the grid is important for two main reasons; namely, the number of elements in the mesh can be kept low, even if the model suffers from areas with extremely fine detail, and secondly, the FDTD timestep can be kept long. Recall the simulation stability condition imposed by the Courant--Friedrichs--Lewy (CFL) condition \cite{cfl, cfl_yee}:

\begin{equation}\label{eq:cfl}
\Delta t \leq \frac{\displaystyle 1}{\displaystyle c\sqrt{\frac{1}{{\Delta x}^2}+\frac{1}{{\Delta y}^2}+\frac{1}{{\Delta z}^2}}},\\
\end{equation}
where $\Delta t$ is the timestep imposed by the constraint, $c$ is the speed of light, and $\Delta x$, $\Delta y$ and $\Delta z$ correspond to the minimum cell edge length in the $x$, $y$ and $z$ axis, respectively. Therefore, by ensuring no cell exceeds a specified fraction, such as $\lambda/300$, tiny details in the geometry of the model---including e.g.,~thin PCB and microstrip traces with a thickness of $\sim35${\ \textmu}m---which fail to contribute anything to the model can be eliminated, as they are electromagnetically unnecessary. This in turn minimizes the total number of timesteps $N_\mathrm{ts}$, ultimately leading to a faster simulation; often with a considerable speedup factor.

\subsection{Optional Parameters}

As mentioned, sudden changes in the material properties between two neighboring mesh cells can have a meaningful impact on the output of a simulation. Let $(\varepsilon_n)$ and $(\mu_n)$ be the sequences of an axis $\in \mathbb{R}^3$, indicating the permittivity and permeability at the $n$-th cell of the axis array, respectively.\footnote{Unlike the sequence describing the position of each mesh line which is strictly increasing, $(\varepsilon_n)$ and $(\mu_n)$ are typically non-injective, as the same material is generally part of several cells along the same axis.} If we consider that $\exists\ m\in\mathbb{N}$, such that

\begin{equation}\label{eq:and_or}
\begin{aligned}
\varepsilon_{m+1} - \varepsilon_m \neq 0\ \ \ &\lor\ \ \ \mu_{m+1} - \mu_m \neq 0\\
\Delta \varepsilon \neq 0\ \ \ &\lor\ \ \ \Delta \mu \neq 0,
\\
\end{aligned}
\end{equation}
it becomes apparent that it is of great importance to ensure the mesh is appropriately refined $\forall m$ for which Eq. (\ref{eq:and_or}) holds true. I.e., for every edge around which such variations are expected, additional lines shall be inserted to assure a smooth transition between materials of different properties. If we assume a material-based boolean operation (addition; shape fuse)\footnote{\url{https://cadquery.readthedocs.io/en/latest/classreference.html\#cadquery.Compound.fuse}} has preceded prior to the geometry being fed to the core segment of the mesh generation algorithm, such that $N_\mathrm{shapes} = N_\mathrm{materials}$, the number of edges and vertices will be minimized. By refining each vertex of the CAD model, we can therefore practically ensure all shape edges are properly refined. Fig.~\ref{fig:sequence} demonstrates edge refinement on a sample microstrip patch antenna case.

By default, edge refinement places 3 additional lines around each side of the corresponding node, symmetrically (yielding a total of 6 additional lines for every $m$). In case the model could benefit from certain special tweaking, this can be optionally adjusted using the \texttt{n} variable, which allows the constant to be configured to a lower or higher value for each plane, individually.

Similarly, the \texttt{res\_fraction} parameter list can adjust the wavelength fraction for each plane ($\lambda/6$ by default), so that finer details can be picked up, or for the mesh to become further adapted to maximize the timestep $\Delta t$ (CFL constraint; Eq. (\ref{eq:cfl})).

\begin{figure*}
  \begin{center}
  \includegraphics[width=0.98\textwidth]{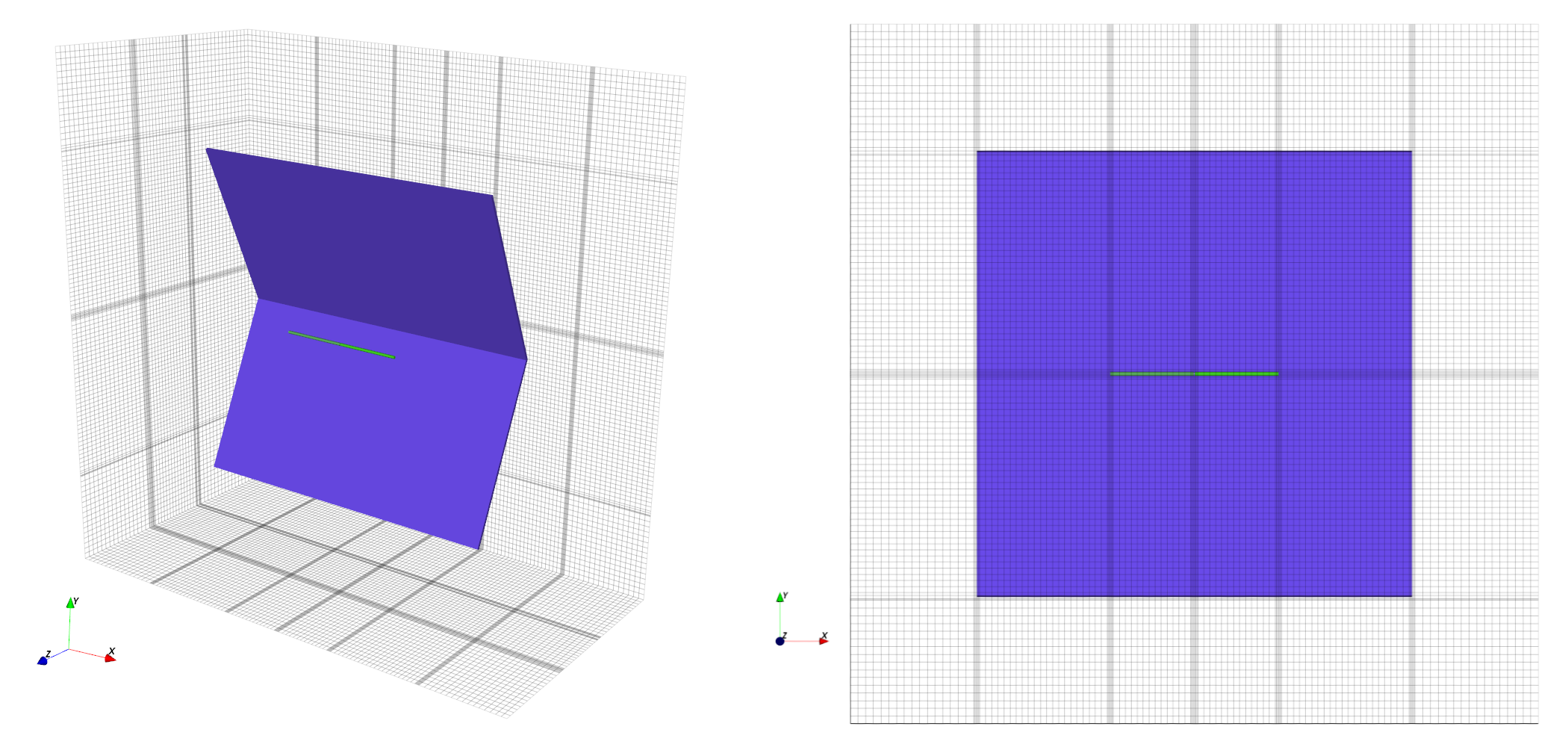}
  \caption{Produced mesh generated using the proposed algorithm on a dipole-fed corner reflector antenna. \textit{Left:} Three-dimensional view of the derived grid. \textit{Right:} Two-dimensional parallel projection of the XY plane, depicting the refined regions and the sparsely-separated grid lines far from the model. The grid consists of $132\times118\times64=996{,}864$ mesh cells and was produced in under 2 seconds on an Intel Core i7-10750H CPU @ 2.60 GHz ($<$1 s for CAD preprocessing and $<$0.3 s for mesh generation).}\label{fig:wire_mesh}
  \end{center}
\end{figure*}

\section{Depth-First Search}

Because the algorithm expects a STEP file as an input (to support a broad range of CAD geometries in a compatible manner), it is important for the individual shapes to be separated from compounds (groups of shapes). However, due to the variety of designs imposing different CAD formats, the expected structure of the relationship between compounds and shapes is unknown. For that reason, we choose to interpret the model's structure as a tree consisting of an arbitrary number of nodes, depending on the input model. The internal nodes represent the compounds, while the leaf nodes represent the shapes of interest we wish to obtain.

While CAD preprocessing is not strictly considered to be a part of the core mesh-generation algorithm, it unquestionably constitutes a critical step required for the derivation of the geometry vertices of the input model. Handling the geometry thus becomes a node searching problem, where a tree data structure needs to be traversed to obtain each individual shape the geometry consists of.

In order to traverse the tree, we can employ the depth-first search (DFS) algorithm \cite{dfs}. Beginning from a root node, DFS works by exploring each branch individually, ultimately retrieving all leaf nodes of interest. This makes DFS an ideal candidate for retrieving the shapes from compounds, as certain compounds may contain several other compounds (arranged in the form of a subtree), before unveiling its shape(s). In other words, nested compounds are possible and expected. Fig.~\ref{fig:dfs} shows an example geometry input, translated into its tree data structure.

\section{Accuracy Evaluation}

\begin{figure*}
  \begin{center}
  \includegraphics[width=\textwidth]{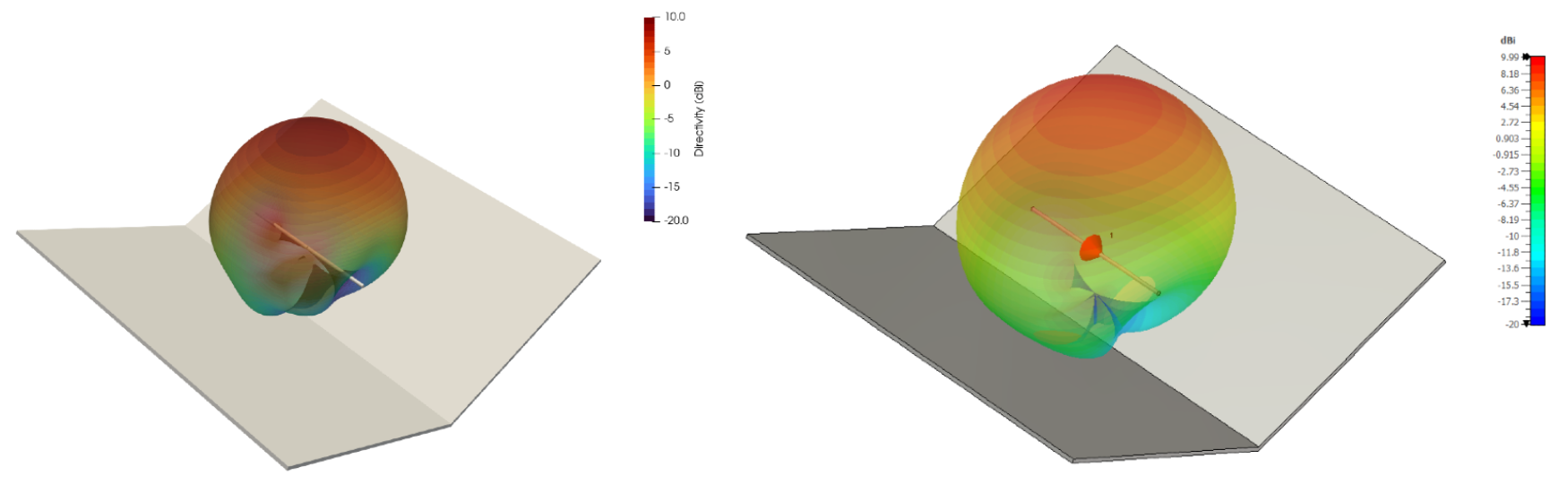}
  \includegraphics[width=\textwidth]{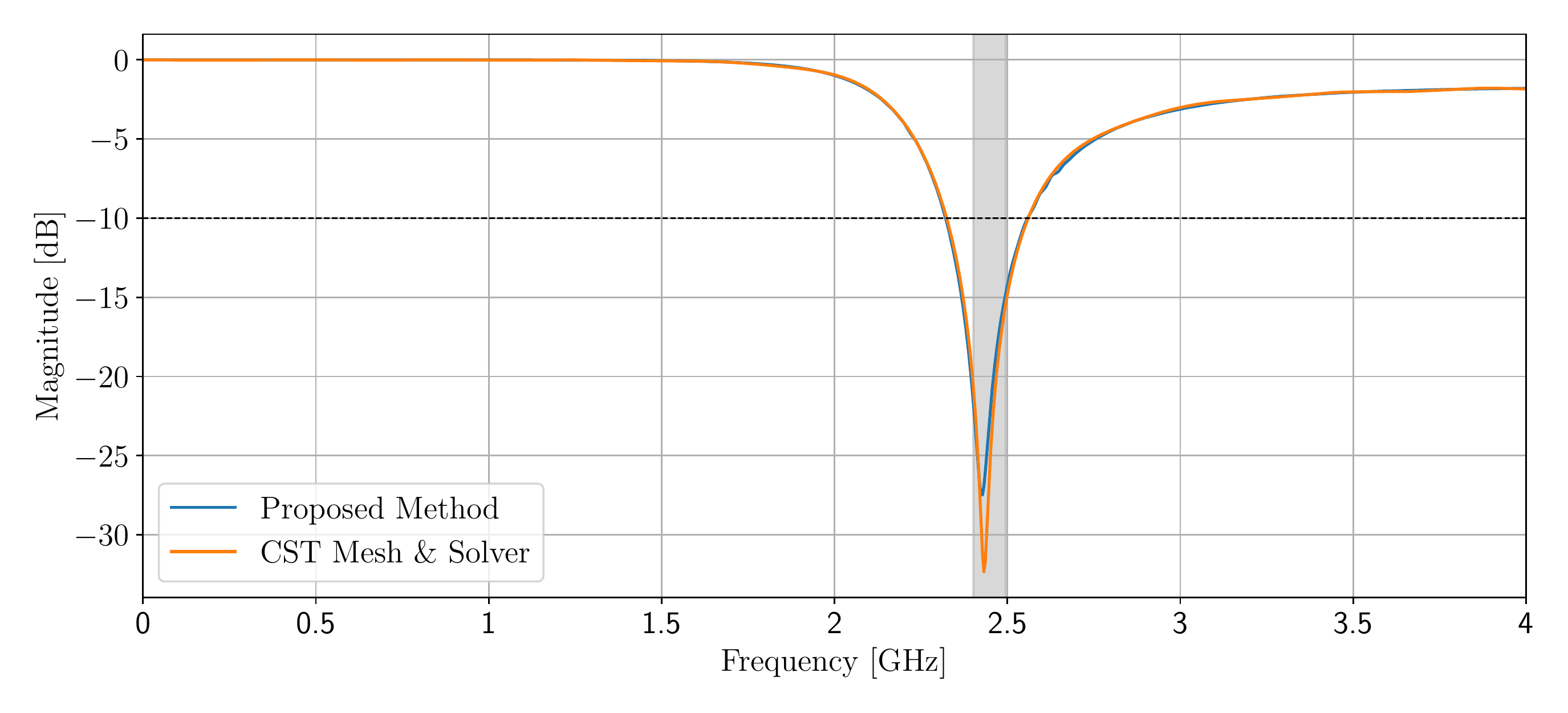}
  \caption{Results derived using the proposed meshing algorithm compared to CST's proprietary mesher and time-domain solver. \textit{Top:} 3D farfield at 2.4 GHz, showing the radiation pattern produced using our method (left) and CST (right). The maximum boresight directivity derived using EC--FDTD and CST's solver is 9.81 and 9.99 dBi, respectively. \textit{Bottom:} Comparison of the reflection coefficient $(|\mathrm{S}_{11}|)$ between the two simulations, showing solid agreement across the entire DC---4 GHz range. The shaded region (2,401---2,495 MHz) corresponds to the 14 channels of the IEEE 802.11b/g standard for wireless local area network (WLAN) communications in the 2.4 GHz band.}\label{fig:wire_results}
  \end{center}
\end{figure*}

In order to evaluate the robustness and quality of the proposed algorithm, the simulation results produced with the generated grids are compared with those of commercial tools. Considering its similarities in terms of mesh generation and time-domain approach to Maxwell's equations, CST Studio Suite has been used as a baseline for our results. In Fig.~\ref{fig:wire_mesh}, we show the mesh produced using the following parameters for a dipole-fed corner reflector antenna designed for Wi-Fi applications:

\begin{itemize}
   \item \texttt{max\_cell\_model} = 40
   \item \texttt{max\_cell\_space} = 30
   \item \texttt{min\_cell\_global} = 300\\
\end{itemize} The simulation has been carried out using the open-source equivalent-circuit finite-difference time-domain (EC--FDTD) solver offered by the openEMS tool \cite{openems, openems_2}. The results (Fig.~\ref{fig:wire_results}) are in agreement with CST in terms of the reflection coefficient, pattern shape, as well as peak directivity.\footnote{The results are also in solid agreement in terms of the time-domain voltage response and $\mathrm{S}_{11}$ phase, but were not deemed meaningfully important to include as separate figures.}

\section{Conclusions}

Despite the numerous advancements in computational electromagnetics over the last decades, no reliable open-source solution had been available for a stage as critical as geometry discretization. While numerous tetrahedral meshing algorithms have existed for many years due to the broad range of applications found in the various fields of simulation engineering, a hexahedral rectilinear mesher---applicable to the popular finite-difference time-domain method---had not been available.

We have presented a robust and automatic approach to tackle this problem in a highly efficient and computationally inexpensive manner, that enables the algorithm to be used by antenna designers and RF engineers to easily mesh their complex geometries in a highly automated and greatly simplified manner. Furthermore, we have compared our results with commercial software, and have demonstrated clear agreement between both the produced grid, as well as the simulation output derived with each mesh and solver.

In future work, we hope to develop an accurate binary classifier based on machine learning, with the aim of introducing a highly intelligent system capable of determining whether a three-dimensional non-uniform structured grid is fine enough, based on mesh convergence. Assuming the training dataset of antenna models with varying mesh resolutions is sufficiently large, such a neural network is expected to drastically reduce the number of mesh cells required to produce an accurate result (using image recognition), leading to a significantly faster simulation. Furthermore, if the false positive rate of the classifier is found to be exceptionally low, conventional mesh convergence analysis could potentially become redundant.

Additionally, considering the openEMS package supports an EC--FDTD implementation in cylindrical coordinates as well, a similar algorithm could be adjusted to be applicable and adaptable to geometries that present curvature, where the FDTD staircase representation problem induced by conventional Cartesian grids could be avoided.


%

\section*{Acknowledgments}

The author would like to acknowledge the help provided by Giorgi Kartvelishvili for the valuable guidance offered towards the setup and testing of solutions related to computational acceleration using compiled code.
\newpage
\ifCLASSOPTIONcaptionsoff
  \newpage
\fi



%

\bibliographystyle{IEEEtran}
\bibliography{IEEEabrv,Bibliography}

\begin{thebibliography}{10}
\providecommand{\url}[1]{#1}
\csname url@rmstyle\endcsname
\providecommand{\newblock}{\relax}
\providecommand{\bibinfo}[2]{#2}
\providecommand\BIBentrySTDinterwordspacing{\spaceskip=0pt\relax}
\providecommand\BIBentryALTinterwordstretchfactor{4}
\providecommand\BIBentryALTinterwordspacing{\spaceskip=\fontdimen2\font plus
\BIBentryALTinterwordstretchfactor\fontdimen3\font minus
  \fontdimen4\font\relax}
\providecommand\BIBforeignlanguage[2]{{%
\expandafter\ifx\csname l@#1\endcsname\relax
\typeout{** WARNING: IEEEtran.bst: No hyphenation pattern has been}%
\typeout{** loaded for the language `#1'. Using the pattern for}%
\typeout{** the default language instead.}%
\else
\language=\csname l@#1\endcsname
\fi
#2}}

\bibitem{davidson2010}
D.~B. Davidson, \emph{Computational Electromagnetics for RF and Microwave
  Engineering}, 2nd~ed.\hskip 1em plus 0.5em minus 0.4em\relax Cambridge
  University Press, 2010.

\bibitem{yee1997}
K.~Yee and J.~Chen, ``The finite-difference time-domain (fdtd) and the
  finite-volume time-domain (fvtd) methods in solving maxwell's equations,''
  \emph{IEEE Transactions on Antennas and Propagation}, vol.~45, no.~3, pp.
  354--363, 1997.

\bibitem{mom}
R.~F. Harrington, \emph{Field computation by moment methods}, ser. IEEE Press
  Series on Electromagnetic Wave Theory.\hskip 1em plus 0.5em minus 0.4em\relax
  Piscataway, NJ: IEEE Publications, Apr. 1993.

\bibitem{fem}
J.~Jin, \emph{\BIBforeignlanguage{en}{The Finite Element Method in
  Electromagnetics}}, 3rd~ed., ser. Wiley - IEEE.\hskip 1em plus 0.5em minus
  0.4em\relax Nashville, TN: John Wiley \& Sons, Mar. 2014.

\bibitem{solver_comparison}
M.~S. Tong and W.~C. Chew, \emph{Computational Electromagnetics}, 2019, pp.
  75--97.

\bibitem{aegmesher}
M.~K. Berens, I.~D. Flintoft, and J.~F. Dawson, ``Structured mesh generation:
  Open-source automatic nonuniform mesh generation for fdtd simulation.''
  \emph{IEEE Antennas and Propagation Magazine}, vol.~58, no.~3, pp. 45--55,
  2016.

\bibitem{gmsh}
\BIBentryALTinterwordspacing
C.~Geuzaine and J.-F. Remacle, ``Gmsh: A 3-d finite element mesh generator with
  built-in pre- and post-processing facilities,'' \emph{International Journal
  for Numerical Methods in Engineering}, vol.~79, no.~11, pp. 1309--1331, 2009.
  [Online]. Available:
  \url{https://onlinelibrary.wiley.com/doi/abs/10.1002/nme.2579}
\BIBentrySTDinterwordspacing

\bibitem{cadquery}
\BIBentryALTinterwordspacing
A.~Urbańczyk, J.~Wright, D.~Cowden, I.~T. Solutions, H.~Y. ÖZDERYA, M.~Boyd,
  B.~Agostini, M.~Greminger, J.~Buchanan, cactrot, huskier, M.~S.
  de~León~Peque, P.~Boin, W.~Saville, B.~Weissinger, Ruben, nopria,
  C.~Osterwood, moeb, A.~Kono, HLevering, W.~Turner, A.~Trhoň, G.~Christoforo,
  just georgeb, A.~Peterson, A.~Grunichev, A.~Gregg-Smith, Bernhard, and
  D.~Anderson, ``Cadquery/cadquery: Cadquery 2.1,'' Feb. 2021. [Online].
  Available: \url{https://doi.org/10.5281/zenodo.4498634}
\BIBentrySTDinterwordspacing

\bibitem{numpy}
\BIBentryALTinterwordspacing
C.~R. Harris, K.~J. Millman, S.~J. van~der Walt, R.~Gommers, P.~Virtanen,
  D.~Cournapeau, E.~Wieser, J.~Taylor, S.~Berg, N.~J. Smith, R.~Kern, M.~Picus,
  S.~Hoyer, M.~H. van Kerkwijk, M.~Brett, A.~Haldane, J.~F. del R{\'{i}}o,
  M.~Wiebe, P.~Peterson, P.~G{\'{e}}rard-Marchant, K.~Sheppard, T.~Reddy,
  W.~Weckesser, H.~Abbasi, C.~Gohlke, and T.~E. Oliphant, ``Array programming
  with {NumPy},'' \emph{Nature}, vol. 585, no. 7825, pp. 357--362, Sept. 2020.
  [Online]. Available: \url{https://doi.org/10.1038/s41586-020-2649-2}
\BIBentrySTDinterwordspacing

\bibitem{cython}
S.~Behnel, R.~Bradshaw, C.~Citro, L.~Dalcin, D.~S. Seljebotn, and K.~Smith,
  ``Cython: The best of both worlds,'' \emph{Computing in Science \&
  Engineering}, vol.~13, no.~2, pp. 31--39, 2011.

\bibitem{numba}
\BIBentryALTinterwordspacing
S.~K. Lam, A.~Pitrou, and S.~Seibert, ``Numba: A llvm-based python jit
  compiler,'' in \emph{Proceedings of the Second Workshop on the LLVM Compiler
  Infrastructure in HPC}, ser. LLVM '15.\hskip 1em plus 0.5em minus 0.4em\relax
  New York, NY, USA: Association for Computing Machinery, 2015. [Online].
  Available: \url{https://doi.org/10.1145/2833157.2833162}
\BIBentrySTDinterwordspacing

\bibitem{cfl}
R.~{Courant}, K.~{Friedrichs}, and H.~{Lewy}, ``{{\"U}ber die partiellen
  Differenzengleichungen der mathematischen Physik},'' \emph{Mathematische
  Annalen}, vol. 100, pp. 32--74, Jan. 1928.

\bibitem{cfl_yee}
K.~{Yee}, ``{Numerical solution of inital boundary value problems involving
  maxwell's equations in isotropic media},'' \emph{IEEE Transactions on
  Antennas and Propagation}, vol.~14, no.~3, pp. 302--307, May 1966.

\bibitem{dfs}
R.~Tarjan, ``Depth first search and linear graph algorithms,'' \emph{SIAM
  JOURNAL ON COMPUTING}, vol.~1, no.~2, 1972.

\bibitem{openems}
\BIBentryALTinterwordspacing
T.~Liebig. openems - open electromagnetic field solver. General and Theoretical
  Electrical Engineering (ATE), University of Duisburg-Essen. [Online].
  Available: \url{https://www.openEMS.de}
\BIBentrySTDinterwordspacing

\bibitem{openems_2}
\BIBentryALTinterwordspacing
T.~Liebig, A.~Rennings, S.~Held, and D.~Erni, ``openems – a free and open
  source equivalent-circuit (ec) fdtd simulation platform supporting
  cylindrical coordinates suitable for the analysis of traveling wave mri
  applications,'' \emph{International Journal of Numerical Modelling:
  Electronic Networks, Devices and Fields}, vol.~26, no.~6, pp. 680--696, 2013.
  [Online]. Available:
  \url{https://onlinelibrary.wiley.com/doi/abs/10.1002/jnm.1875}
\BIBentrySTDinterwordspacing

\end{thebibliography}



%








\end{document}